\documentclass[namedreferences]{solarphysics}
\usepackage[hyperref,optionalrh,solaromanenum]{spr-sola-addons} % For Solar Physics 
\usepackage{epsfig}                     % For eps figures, old commands
\usepackage{graphicx}                    % For eps figures, newer & more powerfull
\usepackage{color}                       % For color text: \color command
\usepackage{breakurl}                         % For breaking URLs easily trough lines
\begin{document}

\begin{article}

\begin{opening}

\title{Muon Excess at Sea Level during the Progress of a
Geomagnetic Storm and High Speed Stream Impact Near  the Time of Eath's Heliospheric Sheet Crossing}
%CHANGE
%Energetic particle injection on Earth from a coronal high speed stream during  the progress  of a geomagnetic storm
%Energetic particle injection on Earth driven by a coronal high speed stream during  the progress  of a geomagnetic storm.

%%%%%%%%%%%%%%%%%%%%%%%%%%%%%%%%%%%%%%%%%%%%%%%%%%%
%% Authors Names
%
% \author[addressref={},corref,email={}]{\inits{}\fnm{}\lnm{}\orcid{}}
\author[addressref={1},corref,email={}]{C. R. A. Augusto}
\author[addressref={1},corref,email={navia@if.uff.br}]{C. E. Navia}
\author[addressref={1},corref,email={}]{M. N. de Oliveira}
\author[addressref={2},corref,email={}]{A. A. Nepomuceno}
\author[addressref={3},corref,email={}]{V. Kopenkin}
\author[addressref={4},corref,email={}]{T. Sinzi}
%%%%%%%%%%%%%%%%%%%%%%%%%%%%%%%%%%%%%%%%%%%%%%%%%%%
%% Runningheads
%
\runningauthor{Augusto et al}
%CHANGE
\runningtitle{Energetic particle injection on Earth driven by a coronal high speed stream}

%%%%%%%%%%%%%%%%%%%%%%%%%%%%%%%%%%%%%%%%%%%%%%%%%%%
%% Affilations 
%% id shold be the same with \author addressref value.
%\address[id={}]{}

\address[id={1}]{Instituto de F\'{\i}sica, Universidade Federal Fluminense, 24210-346,
Niter\'{o}i, RJ, Brazil}

\address[id={2}]{Departamento de Ci\^{e}ncias da Natureza, UFF, Rio das Ostras, RJ, Brazil}

\address[id={3}]{ Research Institute for Science and Engineering, Waseda University, Shinjuku, Tokyo 169, Japan.}

\address[id={4}]{Rikkyo University, Toshima-ku, Tokyo 171, Japan}

%%%%%%%%%%%%%%%%%%%%%%%%%%%%%%%%%%%%%%%%%%%%%%%%%%%
%%% Abstract

\begin{abstract}
In this article we present results of the study on the association between the muon flux variation at ground level, registered by the \textit{\textit{New-Tupi}} muon telescopes ($22^0 53'00''S,\; 43^0 06'13'W$; 3 m above sea level) and a 
geomagnetic storm of 25\,--\,29 August 2015 that has raged for several days as a result of a coronal mass ejection (CME) impact on Earth's magnetosphere.
A sequence of events started with an M3.5 X-ray class flare on 22 August 2015 at 21:19 UTC. The \textit{\textit{New-Tupi}} muon telescopes observed a Forbush decrease (FD) triggered by this geomagnetic storm, with onset on 26 August 2015. 
After the Earth crossed a heliospheric current sheet (HCS), an increase in the particle flux was observed on 28 August 2015 by spacecrafts and ground level detectors.
The observed peak was in temporal coincidence with the impact of a high speed stream (HSS). We study this increase, that has been observed with a significance above 1.5\% by ground level detectors in different rigidity regimes.
We also estimate the lower limit of the energy fluence injected on Earth. 
In addition, we consider the origin of this increase, such as acceleration of particles by shock waves on the front of the HSS and the focusing effect of the HCS crossing. 
Our results show possible evidence of a prolonged energetic (up to GeV energies) particle injection within the Earth atmosphere system, driven by the HSS. 
In most cases these injected particles are directed to polar regions. 
However, the particles from the high energy tail of the spectrum can reach middle latitudes, and that could  have consequences for the atmospheric chemistry, for instance, the creation of NOx species may be enhanced and can lead to increased ozone depletion. 
This topic requires further study.
\end{abstract}

%%%%%%%%%%%%%%%%%%%%%%%%%%%%%%%%%%%%%%%%%%%%%%%%%%%
%% Keywords
%
\keywords{HSS, Geomagnetic Storm, High energy particles injection}

\end{opening}
%-------------------------------------------------
\section{Introduction}

Space weather conditions in Earth's interplanetary environment are constantly influenced  by the transient disturbances in the 
interplanetary medium propagating from the Sun, such as coronal mass ejections (CMEs) and high speed solar wind streams (HSSs). 
After the solar maximum period, the coronal holes are more stables and a HSS can reappear with a $\sim 27$-day period. 
A  persistent stream is called as corotating interaction region (CIR) and it is the region where the fast wind from such holes catches up with slow solar wind, forming shock waves.
The massive bursts of plasma and magnetic fields released into the interplanetary medium are often associated with solar 
flares \citep{tang89,gosling91,gonzalez99}. 

Solar flares occur whenever there is a rapid large-scale change in the Sun's magnetic field \citep{sakurai71}.
Flares are classified into five major classes, X, M, C, B, and A, with X corresponding to the 
\textit{Geostationary Operations Environmental Satellite} (GOES)  energy flux (in the energy band 100-800 pm) 
in excess of $10^4 \,\textrm{Watts} \,\textrm{m}^{-2}$ at Earth \citep{fletcher2011}.
The frequency of occurrence of solar flares varies, following the 11-year solar cycle. 
The current Cycle 24 has already shown evidence of anomalously low solar activity \citep{gopalswamy15}.

The flow of charged particles from the solar corona exhibits a variability over a broad range of time and space.
The solar wind emitted by the regions at high latitudes is fast (500\,--\,800 km s$^{-1}$), at lower latitudes 
the velocities tend to (300\,--\,400 km s$^{-1}$) \citep{phillips95,tsurutani2011}. 
High speed solar wind streams emanate from solar coronal holes, dark regions in the corona where the magnetic field is ``open'' into interplanetary space.
A coronal hole can reappear with the 27-day rotation period of the Sun. 
During solar maximum and high solar activity, coronal holes exist at all latitudes, but are less persistent \citep{tsurutani06}. 

Due to the rotation of the Sun, a fast solar wind stream, typically from a coronal hole, interacts with slow solar wind,  
producing a stream interaction region (SIR).
Streams from persistent coronal holes over multiple rotations form regions of enhanced magnetic field strength and particle 
density that are known as co-rotating interaction regions (CIRs) \citep{smithwolf1976,tsurutani06}.
%CHANGE
CIRs are often coincident (at 1 AU) with an intersection of a major feature of the heliosphere called the heliospheric current sheet (HCS), a 
transition zone that separates regions of opposing heliospheric magnetic field polarity \citep{willcox1965,tsurutani95}.
During 27-day rotation, the HCS passes over Earth a number of times (usually between two and six times) \citep{smith01,thomas2014}.
HCSs can be highly distorted and can have substructures caused by out-flowing transients \citep{villante1979,rouillard2007}.
%%%CHANGE

The charged particles emitted by a flare follow the solar magnetic field lines (Parker spiral), which define the sunward magnetic field direction \citep{parker1963}.
In general, solar flares originated in active areas located in the western sector of the Sun (in the helio longitude) have a good magnetic connection with the Earth.
If the interplanetary counterparts of coronal mass ejections (ICMEs), and CIRs have a significant southward interplanetary magnetic field (IMF) component, 
then after reaching Earth's magnetosphere, they may lead to geomagnetic storms \citep{tsurutani88,gonzalez99,gonzalez94,tsurutani06}.

When the the Earth-directed transient disturbances in the interplanetary medium pass by Earth's magnetosphere, the ground-based
detectors can observe depressions of cosmic-ray intensity, the so-called Forbush decreases \citep{forbush37,forbush38}. 
A magnetic cloud  and its associated bow shock can act as a barrier, preventing galactic cosmic rays (GCR) from reaching the Earth \citep{cane00}.
The temporal decrease in the observed galactic cosmic ray intensity is followed by a gradual recovery that typically lasts up to several days \citep{forbush37,lockwood71}.
The amplitude of a Forbush decrease observed by a particular detector depends on several factors: the size of the CME, the strength of the magnetic
field in the CME, the proximity of the CME to Earth, and the location of the detector on Earth \citep{cliver96,kim08,belov14,heber15}. 

The solar wind carries some of the Sun's magnetic field. 
If an interplanetary high speed stream and the Earth's magnetic field are oriented in opposite directions, an interaction between the two can take place.
This process, in which the two magnetic fields lines interconnect, is known as ``magnetic re-connection''  \citep{dungey61,angelopoulos08,frey03}.
%CHANGE
The reconnection between the solar-wind magnetic field and the Earth's magnetic field enhances the direct magnetic connection between the solar wind and the Earth's magnetosphere.
%CHANGE 
The electrically charged particles from solar wind or high energy particles, accelerated by magnetic shocks at the front edges of high-speed stream, could flow through the shield.
If  reconnection occurs over a long period of time, the energy input to the magnetosphere during a HSS event 
and CIR storms can be comparable to or even greater than the input during dramatic solar flares and large CMEs.

Our experiment is dedicated to studying  transient events of diverse origins and space weather phenomena \citep{augusto12}.
In this article we report the results of an ongoing survey on the association between the muon flux variation at ground level 
(3 m above sea level) registered by the \textit{New-Tupi} telescopes installed at the campus of the Universidade Federal Fluminense, Niteroi city, Rio de Janeiro state, Brazil. 
The physical location is inside the South Atlantic Anomaly (SAA) region.
Originating from the decay of charged pions and kaons produced by galactic primary cosmic rays in the atmosphere, the atmospheric muon flux is the main source of background at sea level and in the energy region of sub-GeV to GeV \citep{gaisser2004}. 
The pressure variation influence on the muon counting rate at sea level is weak.
The barometric coefficient for muons is at least seven times lower than the barometric coefficient for protons and neutrons. 

In this article we analyze the geomagnetic storm with onset on 25 August 2015 reaching G2 (moderate) levels that has raged for several days as a result of the 
combined influence of a CME and a coronal hole HSS.
We also report an observation of an increase in the counting rate of detectors on 28 August  2015 16:40 UT in temporal coincidence with a high speed stream.
The narrow peak was osberved after the Earth crossed a heliospheric current sheet.
The \textit{New-Tupi} muon telescopes observations are studied in correlation with the data obtained by other ground based experiments and space-borne detectors.

This article is organized as follows. 
In Section 2, we give a brief description of the \textit{New-Tupi} telescope. In Section 3 experimental data sets and methodology are presented.
Section 4 presents experimental observations and the results of the association between muon variations at ground level and data reported by 
other ground based experiments and space-borne detectors during the geomagnetic storm of 25\,--\,29 August 2015.
We found a correlation between the observed time of the muon excess (positive or negative) signal on Earth (the \textit{New-Tupi} telescopes) 
with the trigger time of the interplanetary disturbances registered by the satellites located at Lagrange point L1 
using data from the Advance Compositon Explorer (ACE) and at a geosynchronous orbit (GOES).
In particular, we focused on the observation of the FD on 25 August 2015 and the  enhancement in the counting rate of ground detectors on 28 August 2015. In Section 5 we explore several interpretations of this signal.
Section 6 is devoted to the analysis of the recovery time of the Forbush decrease, the estimation of the integrated time fluence of the muon enhancement, and the rigidity dependence of the FD and muon enhancement. In Section 7 we present our conclusions.

\section{The \textit{New-Tupi} Muon Telescope}

The purpose of the \textit{New-Tupi} muon telescopes is to learn more about the relationship between the transient interplanetary disturbances propagating from the Sun  and the response of the Earth`s space environment (space weather).
The \textit{Tupi} experimental apparatus  registers the muon intensity in the atmosphere initiated by cosmic rays (mainly protons).
The \textit{New-Tupi} muon telescopes (Niteroi, Brazil,  22.9$^0$ S, 43.2$^0$ W, 3 m above sea level) are within the central 
region of the South Atlantic Anomaly (SAA)  and close to its center 26$^0$ S, 53$^0$ W,  where the geomagnetic field intensity is the lowest Earth.
This is the region where the inner Van Allen radiation belt makes its closest approach to the Earth's surface. 
As a consequence of this behavior, the SAA region is characterized by an anomalously weak geomagnetic field strength (less than 28,000 nT) \citep{barton97}. 
Thus, the shielding effect of the magnetosphere is not quite spherical but has a hole in this region.

\begin{figure}
\vspace{-1.0cm}
\centerline{\includegraphics[width=0.95\textwidth,clip=]{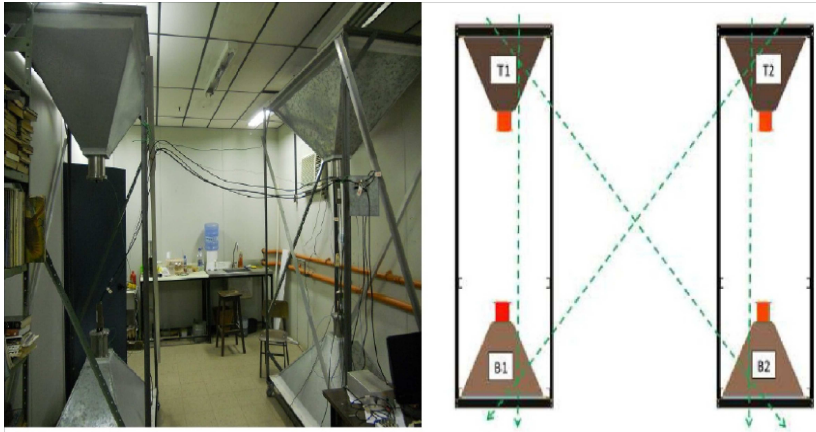}}
\vspace{-7.0cm}
\caption{Left: Photograph of the \textit{New-Tupi} telescope. Right: General scheme of the \textit{New-Tupi} telescope.  This configuration allows to measure the muon flux from three directions, the vertical (zenith), west and east (with an inclination of 45 degrees to the vertical, as is indicated by the dash lines).}
\label{fig1}
\end{figure}

The \textit{New-Tupi} muon telescope consists of four identical particle detectors.
Each detector is constructed on the basis of a commercial plastic scintillator (Eljen EJ-208)  with size (150cm x 75cm x 5cm)  and a photomultiplier (Hamamatsu R877) of 127 millimeters in diameter, and placed inside a box with a truncated square pyramid shape with four slopes. 
The detector assembly is connected to the Ortec ScintiPack$^{TM}$ photomultiplier base (Model 296).
The photomultiplier base includes high voltage divider, amplifier, and high voltage power supplier. 
The data acquisition system is made on the basis of a National Instrument (NI) data acquisition (DAQ) device with eight analog input (AI) channels, working at a rate of 1 MHz.
Each detector registers signals above the threshold value, corresponding to an energy of $\sim$ 100 MeV
deposited by particles that reach it.
The output raw data are stored at a rate of 1 Hz. 

Figure 1 (left) shows a photography of the \textit{New-Tupi} telescope. 
The four detectors are placed in pairs, with the detectors T1 and T2 at the top, and B1 and B2 at the bottom, as shown in Figure 1 (right).
This layout allows to measure the muon flux from three directions, the vertical (zenith), west and east (with an inclination of 45 degrees).
The telescopes register the coincidence rate for the  vertical incidence using the pairs of detectors (T1,B1) and (T2,B2), as well as the cross coincidences between T1 and B2 (west incidence) and the T2 and B1 (east incidence).
The separation between the detectors (vertical and horizontal)  is 2.83 m.

\section{Data Sets and Methodology}

We compare our observations with publicly available data obtained by space-borne detectors: 
ACE and GOES (\url{http://www.swpc.noaa.gov} and \url{http://sohowww.nascom.nasa.gov}). 
ACE is located in the L1 Lagrangian point and provides continuous real-time monitoring of space weather and solar wind information. 
The GOES satellites circle the Earth in a geosynchronous orbit (\url{http://www.oso.noaa.gov/goes/index.htm}) and provide data on the solar
X-ray emissions, the magnitude and direction of Earth's ambient magnetic field, as well as solar proton flux (\url{http://www.goes.noaa.gov}).
We also use data obtained by ground-based neutron monitors at at Mexico, Apatity, Thule stations, using the real 
time Neutron Monitor Database (NMDB) (\url{http://www.nmdb.eu}).

Identification of interplanetary structures and associated solar activity was based on the nomenclature 
and definitions given by the satellite observations
To reflect global information about current magnetosphere activity, we use  the most widely used indicators, 
such as the estimated 3-hour planetary Kp index  (it  runs from 0 to 9) and the ring current Dst \citep{bartels63,sugiura64,rostoker72,mayaud80}.
The interval of large decrease of Dst indicates a geomagnetic storm \citep{gonzalez87,gonzalez94}.
Kp indices of 5 or greater indicate storm-level geomagnetic activity.
The estimated 3-hour planetary Kp index is derived at the NOAA Space Weather Prediction Center using data from the ground-based magnetometers
(\url{http://www.swpc.noaa.gov/rt_plots/kp_3d.html}).
The Dst index is presently compiled by the World Data Center for Geomagnetism at Kyoto, Japan (\url{http://wdc.kugi.kyoto-u.ac.jp}).
A negative Dst value means that Earth's magnetic field is weakened. 
The G scale used by the NOAA National Weather Service is based on the planetary K index and measure the geomagnetic 
effects on physical infrastructure \citep{molinski2000}.
%http://www.aurora-service.eu/aurora-school/all-about-the-kp-index/
A magnetic perturbation that reaches the category of G1 (minor level) is sufficient to enhance the direct magnetic 
connection between the solar wind and the Earth's magnetosphere, resulting in the particle flow through the region \citep{larsen07,ostgaard07}. 

The significance level of the signal in the \textit{Tupi} telescope is defined as ${S}_i=\frac{(C^{(i)}-B)}{B}$,  
where $C^{(i)}$ is the measured number of counts for the signal in the bin $i$ and $B$ is the average background count. 
The temporal structure of the signal (rise, fall, step-like jumps, peaks, \textit{etc.}) is used for further identification.
In order to develop a more specific analysis, we also apply the continuous wavelet  transform (CWT), looking at the time-frequency map information \citep{torrence98,rybak01}.

\section{Experimental Observations}

A geomagnetic storm reaching the condition of G2 (moderate) level ($Kp=6$) was observed on 25 August 2015 following a disturbance in Earth's 
magnetosphere due the transient effects of the CME.
The events started to unravel with an M3.5 X-ray class flare on 22 August 2015 at 21:19 UT 
in the active region (AR) 2403 (\url{http://www.solarmonitor.org}) located at that time at N11W76 (solar disk coordinates), on the western limb. 
The solar flare was accompanied by a CME.
The travel time of the CME was estimated as $ \sim 71$ hours. 

\begin{figure}
\vspace{+0.0cm}
\centerline{\includegraphics[width=1.00\textwidth,clip=]{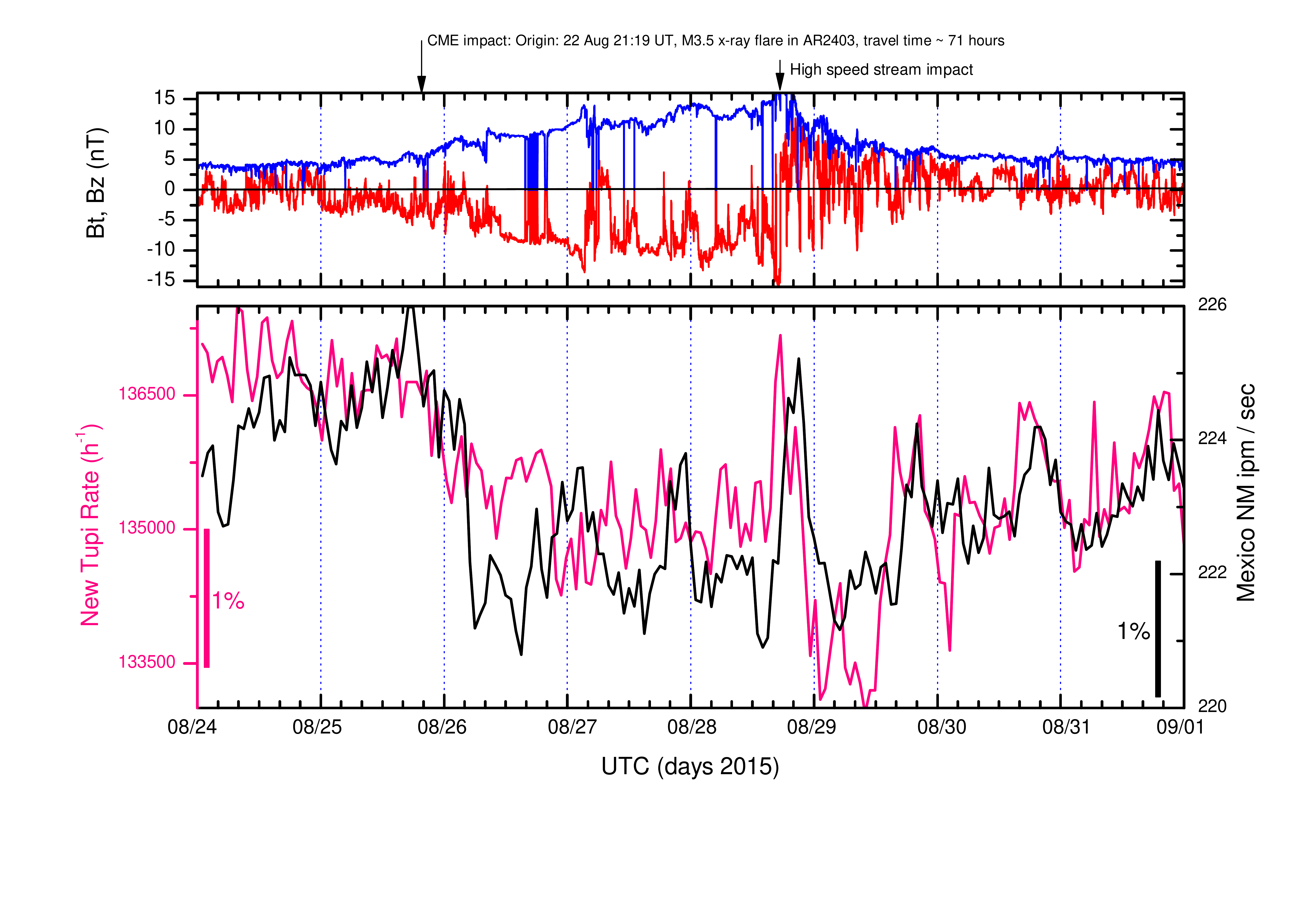}}
\vspace{-1.0cm}
\caption{Top panel: the solar wind  magnetic field components $B_{\rm t}$ and $B_{\rm z}$ observed by the MAG instrument 
\cite{smith98} onboard the ACE satellite located at the Lagrange point L1 on 24\,--\,31 August 2015. Bottom panel: the hourly counting rate in the vertical \textit{New-Tupi} telescope and the counting rates (impulse per second) of the cosmic-ray nucleonic component obtained from the neutron monitor installed in Mexico City.}
\label{fig2}
\end{figure}

From 25 August 2015  the solar wind magnetic field component $B_{\rm z}$ started to be negative.
%CHANGE
The $B_z$ component of the interplanetary magnetic field is measured in geocentric solar magnetospheric (GSM) coordinates, so
the $x$-axis $B_{\rm x}$ points directly towards the Sun, and the $z$-axis points northwards and lies along the magnetic axis of the Earth.
Figure 2 shows a comparison between the $B_{\rm t}$ and $B_{\rm z}$ components (top panel) observed by the ACE spacecraft at the Lagrange point 
L1 (approximately $1.5 × 10^6$ km ), the counting rates (impulse \textit{per} second) of the cosmic-ray nucleonic component 
obtained from the neutron monitor installed in Mexico City (19.33$^{0}$ N, 260.80$^{0}$ E;  2274 m 
above sea level, geomagnetic rigidity 8.2 GV), and the 30 min muon counting rate in the vertical  \textit{New-Tupi} telescope (bottom panel).
It is possible to note that the $B_{\rm z}$ component gradually decreases during 25\,--\,26 August 2015 due to the impact of the interplanetary 
CME reaching the magnetosphere. 
There is a corresponding fall in the counting rate in the detectors at ground level.
The ground telescopes observed a Forbush decrease triggered by this geomagnetic storm.
%%%% CHANGE
As can be seen in Figure 2, the Mexico data points lag \textit{New-Tupi}.
While the Stormer rigidity cutoff  in the dipole approximation of the geomagnetic field is $\sim 9.2$ GV in the \textit{New-Tupi} location, 
the total geomagnetic field ($\sim 23$ mT) and  the magnetic horizontal component ($B_{\rm h}$ $\approx$ 18 mT) are low.
Thus, if we consider the vertical incidence particles, then the sensitivity of the detector located at the \textit{New-Tupi} site is 
close to the detector located at the South Pole.

Geomagnetic activity reached a G2 moderate level on 27 August 2015 and persisted at these level through 28 August 2015 due to the 
slow-moving 22 August 2015 CME.
Figure 3 shows a comparison between the muon counting rate and the geomagnetic indices: 
the planetary Kp index (top panel), the counting rate in the vertical \textit{New-Tupi} telescope (central panel),
and the Dst index variation (bottom panel). 

\begin{figure}
\centerline{\includegraphics[width=1.00\textwidth,clip=]{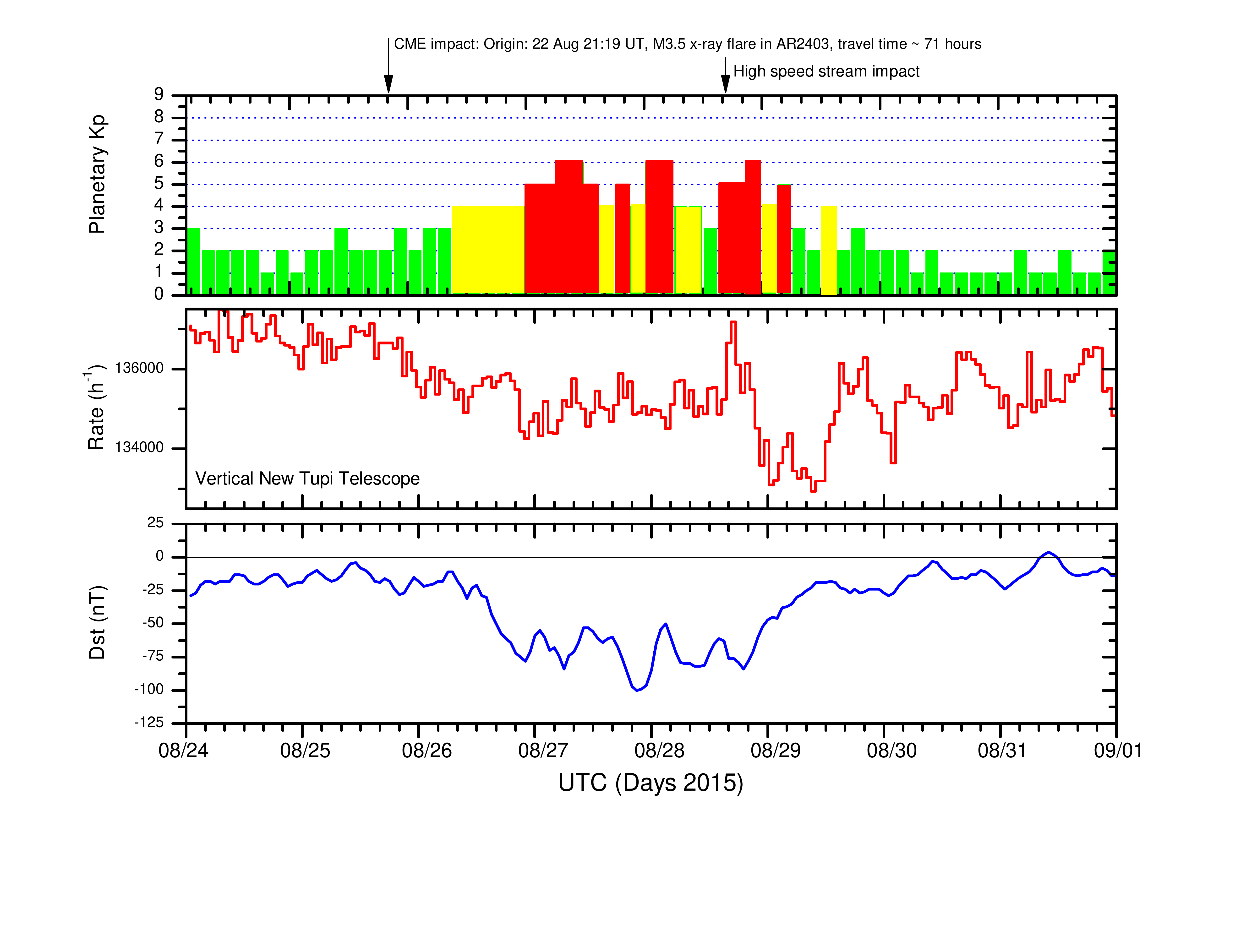}}
\vspace{-1.0cm}
\caption{Top panel: the estimated 3-hour planetary Kp index on 24\,--\,31 August 2015. Central panel: the 1-hour binning muon counting rate in the vertical \textit{New-Tupi} telescope. Bottom panel: the disturbance storm index Dst (hourly values).}
\label{fig3}
\end{figure}

%A fast decrease (oscillation) of the solar wind magnetic field Bz is shown in Figure 4 (top panel).
A sudden increase in the particle flux was observed on 28 August 2015  at $\sim 16:40$ UT by spacecrafts and ground level detectors in temporal coincidence with the impact of a high speed stream (HSS). 
The signal was observed almost simultaneously by ground based detectors in both hemispheres and several continents.
Figure 4 shows a comparison between the $B_{\rm t}$ and $B_{\rm z}$ solar wind magnetic component observed by the ACE satellite (top
panel), the 5 min averaged integral  proton flux with energies above 1 MeV observed by the GOES satellite,  the counting rates 
(30-min resolution) in the \textit{New-Tupi} vertical telescope, and the counting rates (impulse \textit{per} second) of 
the cosmic-ray nucleonic component obtained by the  Apatity  (67.57$^0$ N, 33.40$^0$ E; 181 m above sea level, geomagnetic rigidity 0.65 GV) 
and Thule (76.5$^0$ N, 68.70$^0$ E; 26 m above sea level, geomagnetic rigidity 03 GV) neutron monitors.
The origin of this excess is discussed in the next section.

\begin{figure}
\centerline{\includegraphics[width=1.00\textwidth,clip=]{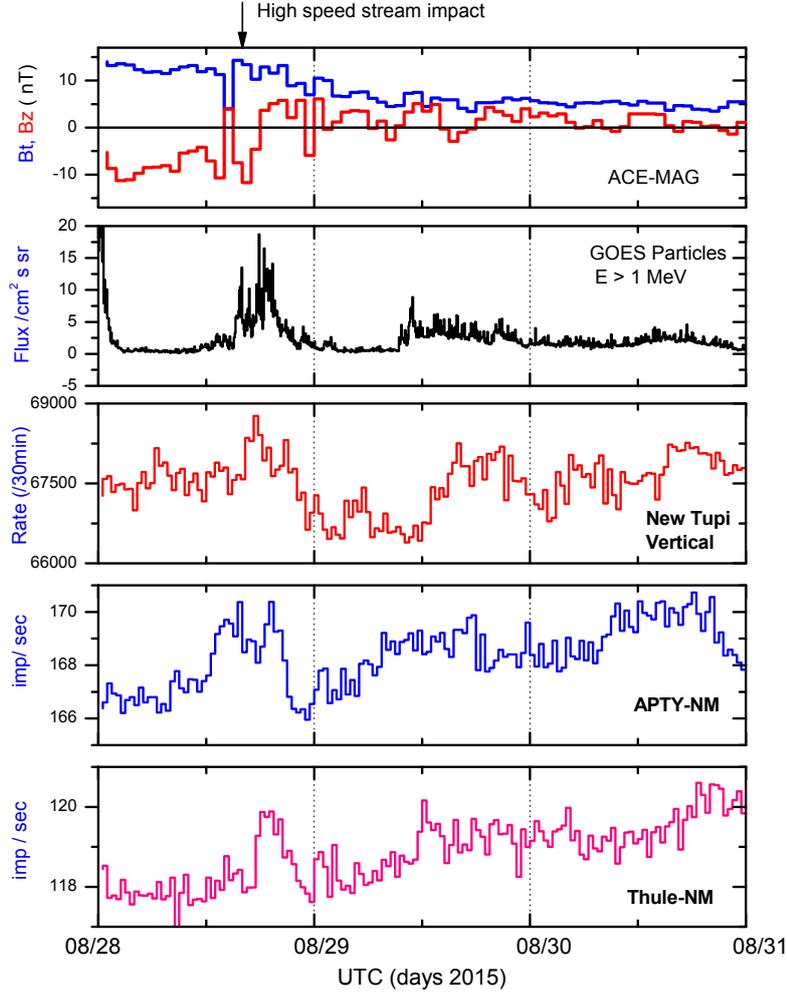}}
\vspace{-2cm}
\caption{Top panel: the solar wind magnetic field components $B_{\rm z}$ and $B_{\rm t}$ observed by the ACE satellite on 
28\,--\,30 August 2015. Second panel: the 5-min averaged integral  proton flux with energies above 1 MeV observed by the GOES satellite. 
Third panel: the 30-min muon counting rate in the vertical \textit{New-Tupi} muon telescope. Fourth and fifth panels: the counting rates (impulse per second) of the cosmic-ray nucleonic component obtained from the Apatity and Thule neutron monitors.}
\label{fig4}
\end{figure}

\section{What is the Origin of the 28 August 2015 Narrow Peak ?}

The mechanism that can produce the observed peak on 28 August 2015 has to explain the features and the circumstances of the observations, 
the onset time of the signal, and the magnitude of the excess.
The observation of an increase in the counting rate at ground level detectors requires primary particles above GeV energies
This strongly suggests that the excess was  due to galactic cosmic rays. 
However, this hypothesis, also requires a mechanism to explain how the cosmic rays were injected.

\subsection{Effect of the HCS crossing}

The galactic cosmic rays can travel towards the Sun better in the coherent magnetic-field structure of the HCS \citep{thomas2014}. 
Typically, the main signature of the crossing is an increase of the particle density in the Earth's environment, including an 
increase in the neutron monitor (NM) counting rates at ground level.
This is due to the increased field in the compressed region that acts as a barrier to the galactic cosmic rays propagation, 
called snow plough effect \citep{richardson2004,thomas2014}.

Using the WSA-Enlil model (\url{https://www.ngdc.noaa.gov/enlil_data/2015/}), that is a large-scale, 
physics-based prediction model of the heliosphere, we show in Figure 5 that on 28 August 2015, 
the Earth (marked as the green circle) has already crossed through a fold in the HCS.

\begin{figure}
\vspace{-1.0cm}
\centerline{\includegraphics[width=1.20\textwidth,clip=]{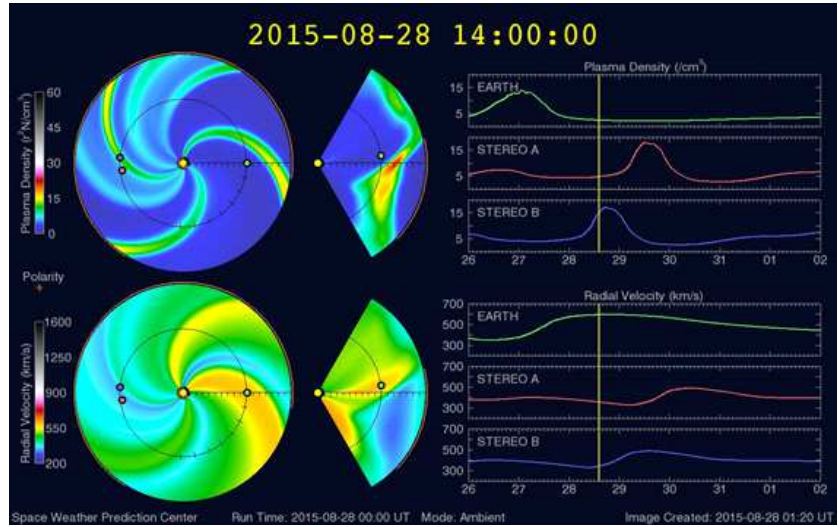}}
\vspace{-10.0cm}
\caption{The WSA-Enlil model shows density (top) and velocity (bottom) of the solar wind on 28 August 2015. 
The Earth (marked as the green circle) has already crossed the HCS.
The Sun is shown by the yellow circle, and the red and blue circles are the twin STEREO spacecrafts in $\sim 1$ AU orbits around the Sun. 
Time is expressed in universal time (UT). Credit: NOAA/SWPC WSA-Enlil model.}\label{fig5}
\end{figure}

Figure 6  shows a comparison between the 1-hour binning muon counting rate in the \textit{New-Tupi} telescope (top panel) and the solar wind parameters 
(density and speed) observed by the ACE satellite observed at L1 (center and bottom panels) in the period from 25\,--\,31 August 2015. 
The shaded area indicates the approximate HCS crossing time and the vertical arrow shows the arrival of a shock associated with a high speed 
stream on 28 August 28 2015.
As can be seen in Figure 6, there is an increase in the particle density observed by ACE at the time of the HCS crossing.
We do not see a clear signal at \textit{New-Tupi} on 26 August 215, apparently due to the arrival to Earth of a magnetic cloud, triggering a geomagnetic storm.
After the lingering CME effects, on 28 August 215 ACE observed another peak in the particle density, and the solar wind velocity showed 
a positive jump (a tyipcal signature of a shock wave).

\begin{figure}
\centerline{\includegraphics[width=1.00\textwidth,clip=]{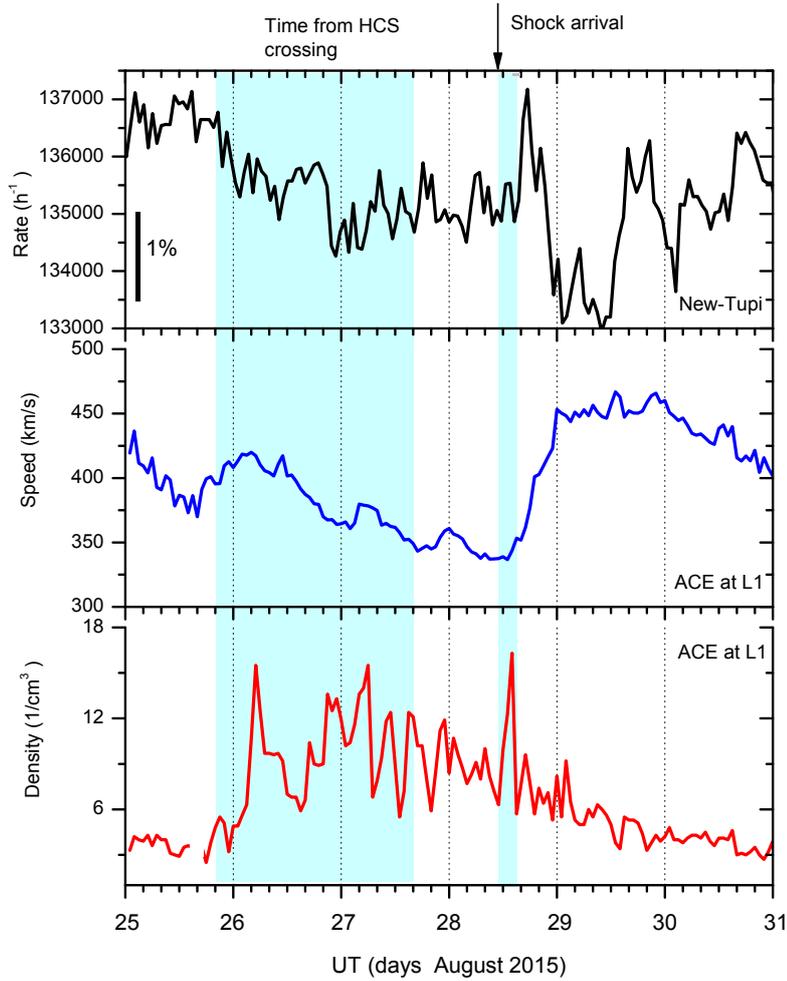}}
\vspace{-2.5cm}
\caption{Comparison between the 1-hour binning muon counting rate in the \textit{New-Tupi} 
telescope (top panel) and the solar wind parameters (density and speed) observed by the ACE satellite (center and bottom panels) 
in the period from 25 August until 31 August. The shaded area indicates the HCS crossing time and the vertical arrow shows the arrival 
of a shock associated with a high speed stream on 28 August.}\label{fig6}
\end{figure}

It is important to note that there are specific cases when a significant enhancement in GCR flux can be observed after the crossing as well \citep{thomas2014}.
Further work is required to quantitatively model and  determine the exact mechanism and specific combination of circumstances which would allow the galactic cosmic rays easily access magnetic field lines and produce the observed excess due to the HCS crossing.

\subsection{Shock Wave Acceleration Hypothesis}

Several signals were observed in association with the impact of the high speed stream on 28 August 2015: the Kp index (Figure 3) shows
almost simultaneous changes in the counting rate at ground detectors in the southern and northern hemispheres (Figure 4),  
and an increase of particles with energies above 1 MeV observed by GOES (Figure 4), and a peak in the particle density observed by ACE (Figure 6).
The ground signal was delayed (about 3 hours) in relation to the satellite trigger (ACE, Lagrange point L1), 
that is a signature of a forward shock wave.
While the amplitude of the GOES excess in the flux level  is somewhat typical of the Earth's radiation belt at 
geosynchronous orbit, we noticed that the observed excess was in temporal coincidence with the others. 
At present this is an empirical observation and future studies of similar associations will be interesting.
The GOES proton flux is typically linked to the solar transient events, and is usually used to classify the radiation storm event
The observation of the 28 August increase in the counting rate at ground level detectors requires primary particles above 
GeV energies and a mechanism to explain how the particles were injected.
In the event analyzed in this article, the excess observed at ground level means that the maximum energies were up to the GeV scale.
Shock acceleration is probably the most widely used particle acceleration mechanism in astrophysics and space physics.
Multiple shocks can be present simultaneously in the solar wind, producing a multiplicity of time scales and determining maximum 
particle energy scales that vary from event to event.
They are in the tail of the high energy spectrum and only the ground-based installations can provide the sufficient geometric 
factor to measure the low intensities of particles accelerated by shocks in the highest energy range, through air showers.
As the \textit{New-Tupi} detector is located at sea level, it is expected that the excess signal  is mainly composed of muons originated by ions, 
that is, ions accelerated by shock waves at the front edges of the transition zones between fast and slow moving solar wind streams. 
In most cases, the injection of supersonic solar wind particles, such as the electrons are commonly directed in the polar regions, 
forming beautiful auroras and the nucleonic component of the solar wind contributes to the creation of NO$_{\rm x}$ species. 
However, high-energy particles can also reach middle latitudes, incrementing the formation of NO$_{\rm x}$. 
The processes NO + O$_{3}$ $\Rightarrow$ NO$_{2}$ + O$_{2}$ and NO$_{2}$ + O $\Rightarrow$ NO+O$_{2}$ 
have consequences for mesospheric chemistry, like a temporal depletion of the ozone layer, making the 
interactions caused by the passage of high speed streams an important topic to understand the space weather effect on Earth \citep{seppala07,kavanagh07}.

\section{Analysis of the 26 August FD Signal and the Narrow Peak
of 28 August 2015}

As one can notice, the confidence level of the FD signal and the narrow peak is not high ($\sim 1\%$).
Here we analyze  both (the \textit{New-Tupi} telescopes and Mexico NM) signals using the continuous wavelet  transform (CWT), looking at the time-frequency map information \citep{torrence98,rybak01}. 
The Fourier spectrum of a signal represents the frequency content of the signal, the signal itself is in the time domain. 
In time-frequency maps, the frequency spectrum is given for each time step so that one can see the evolution of the frequencies. 
The CWT spectra were obtained for the combined data between \textit{New-Tupi} and Mexico-NM and is presented in Figure 5 (bottom panel).  
In this case both signals show high power variance. 
High power (variance) is indicated by red color whereas low power is indicated by blue color. 
One can see that the time intervals with the FD and the narrow peak (most noticeable in color gradient areas) are correctly 
identified and correlate with the time series of the counting rate observed by the \textit{New-Tupi} 
muon telescope and Mexico neutron monitor on 24\,--\,31 August  2015 (top panel). 

\begin{figure}
\vspace{-1.0cm}
\centerline{\includegraphics[width=1.00\textwidth,clip=]{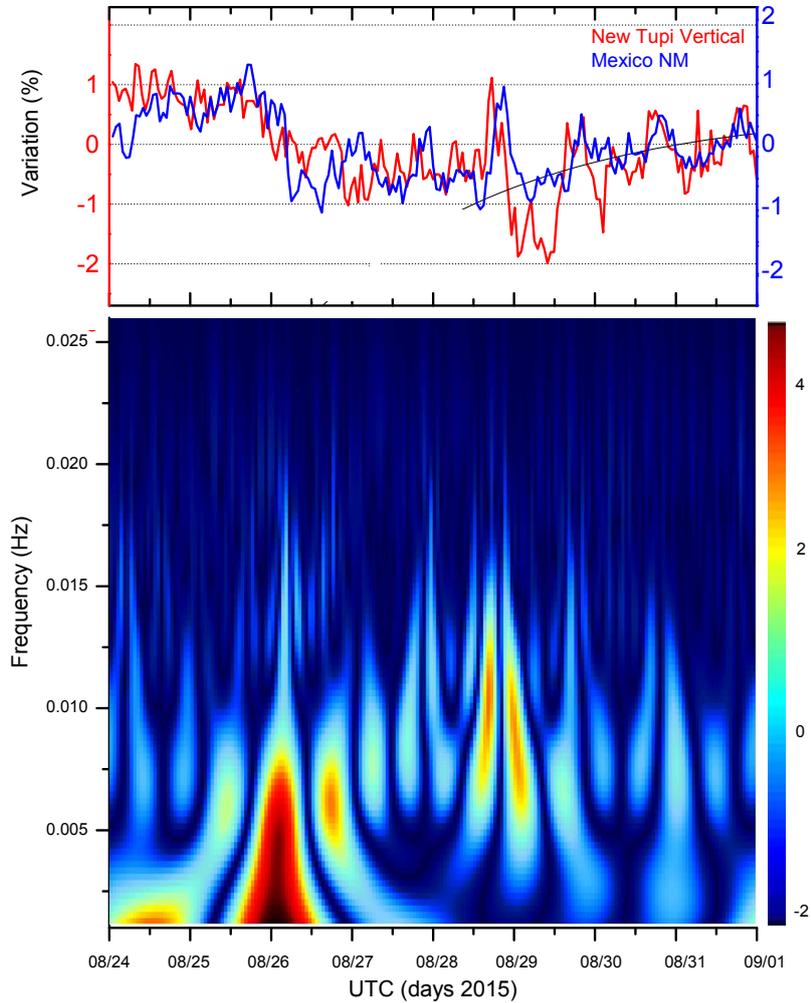}}
\vspace{-1.0cm}
\caption{Top panel: variations (in percents) of the counting rate observed by the \textit{New-Tupi} muon telescope and Mexico neutron monitor on 
24\,--\,31 August  2015. To estimate the recovery time of the Forbush decrease, we use  a nonlinear fit of the data to a single exponent (black curve). Bottom panel: A time-frequency representation of the combined (the \textit{New-Tupi} telescopes and Mexico NM) data for the same period using the continuous wavelet transform (CWT) analysis. The vertical color bar indicates the relative power (variance). High power is indicated by warm (red) colors, whereas low power is indicated by cold  (blue) colors. The intervals corresponding to the fall (26 August 2015)  and the narrow peak (28 August 2015) observed in the counting rate can be easily identified.}\label{fig7}
\end{figure}

\subsection{Recovery time of the Forbush Decrease}

The FD recovery time interval can be defined as the time required for the counting rate to return from the maximum depression to the pre-FD level.
Typically, the shape of the recovery phase of a FD \citep{usoskin08}, especially in case of Forbush events caused by CMEs, can be approximated by an exponential recovery function  $I=I_0-A \exp[-t/\tau]$, where A is an amplitude and $\tau$ is the characteristic recovery time. 
In the present case the situation is more complex due to several factors described above.
At least for two days after the FD of 25\,--\,26 August 2015, the counting rate does not change significantly and can be described by a nearly flat function.
The total duration (from the onset time until the recovery time) has been stretched  (see Figure 2 and Figure 4, top panels).
In addition to that, there is a sudden increase in the counting rate of ground detectors.  
Figure 7 (top panel) summarizes the situation, where the black curve ($\sim 70$ hours after the FD onset) shows a non-linear fit of our data.
The recovery time is estimated here as $\tau=(62 \pm 6)$ hour.

\subsection{Estimation of the Integrated Time Fluence}

On 28 August 2015 at $\sim 16:40$ UT (13:40 local time), the  field of view of the vertical telescope was near the direction of the high 
speed stream impact.
In order to obtain information of the energy injected in the atmosphere during the counting rate enhancement on 
28 August 2015, the integrated time fluence has been derived here. 
The first task is to estimate the vertical muon flux background, using the relation flux = rate/G, 
where rate is the counting rate of the vertical telescope and G is the geometric factor of the vertical telescope, 
G = 6700 cm$^2$ sr \citep{sullivan71}. 

We have used the counting rate on 27 August 2015 as the counting rate background.
Taking into account that the counting rate efficiency of the telescope is about 95\%, the background vertical flux is obtained as

\begin{equation}
I_0=\frac{\rm rate}{\rm G}=(5.9 \pm 0.7) \times 10^{-3} \rm (cm^2 s sr)^{-1}.
\end{equation}
This value is in agreement with the vertical muon flux, obtained at sea level at almost equivalent places \citep{pal12}.

The second task is to obtain the excess of the signal over the background, taking into account that the 
duration of the counting rate enhancement was $T=5.82$ hour. 
The average counting rate in this period (signal (S) plus background (B)) was  $C_{S+B}=(37.8 \pm 4.5)$ Hz.
We have also evaluated the background counting rate, using the 24 hour period (27 August 2015).
The average counting rate background  was $C_B=(37.5 \pm 4.5$) Hz.  
Thus, the counting rate variation due to the enhancement is
\begin{equation}
\frac{C_{S+B}-C_B}{C_B}=\frac{0.25}{37.53}=7.4 \times 10^{-3}.
\end{equation}
In similar form we can write
\begin{equation}
\frac{\rm flux(Signal)}{\rm flux(Background)}=\frac{I_{signal}}{5.9 \times 10^{-3} (\rm cm^{-2} s^{-1} sr^{-1})}=7.4 \times 10^{-3},
\end{equation}
giving a vertical flux (signal) linked with the high speed stream impact of 
\begin{equation}
I_{signal}(E \geq 0.1 \rm GeV)=(4.4\pm 0.5) \times 10^{-5} (\rm cm^{-2} s^{-1} sr^{-1}).
\end{equation}

However, the muon flux goes beyond the vertical direction. 
As a good approximation, the angular dependence of the muon flux with the zenith angle, $\theta$, is cosidered as 
$I(\theta) = I(\theta=0) \textrm{cos}^n(\theta)$, with, $n\sim 2$, for muons in the MeV to GeV range,then we integrate over the solid angle around the vertical direction and obtain
\begin{eqnarray}
N_{\mu}&=&I_{Signal} \int_{\Omega} \cos^n(\theta)d\Omega = 2\pi I_0\int_{\theta_1}^{\theta_2}\cos^n(\theta) \sin (\theta) d\theta\nonumber\\
&=&\frac{2\pi I_0}{n+1} \left(\cos^{n+1}(\theta 1)-\cos^{n+1}(\theta 2)\right),
\end{eqnarray}

\noindent
where $\theta_1 = 0^0$ and $\theta_2 = 60^0$. 
The  $\theta_2$ angle defines the effective angular aperture. At  the large zenith angles ($\theta > 60^0$) the Earth's curvature must be considered
This is due to absorption effects in the thicker atmosphere at large zenith angles.
	Thus, the muon flux associated with the observed excess, can be estimated as
\begin{equation}
N_{\mu}= (8.1 \pm 1.0)\times 10^{-5} (\rm cm^{-2}s^{-1}).
\end{equation}

The third task is to estimate the integrated time fluence observed at ground level, it can be obtained  as
\begin{eqnarray}
F_S(x,t)&=&N_{\mu} \times E_{thr} \times T\nonumber\\ 
&=&(9.2\pm 1.1)\times 10^{-2} \rm GeV/cm^2=(1.4\pm 0.2) \times 10^{-4} \rm erg/cm^2,
\end{eqnarray}
where $E_{thr}=0.1 \rm GeV$ is the energy threshold, and $T=5.82$ hour is the duration of the counting rate enhancement. 
We would like to point out that this value of the observed fluence is only the lower limit.
The fluence of the injected high energy particles at top of the atmosphere must be higher than this values. 

\subsection{Rigidity Dependence}

The response of the Earth's magnetosphere to the interplanetary disturbances can be studied with the use of neutron monitors, 
by analyzing the temporal variations of neutron counts at different vertical cutoff rigidities. 
In most cases the observed FD amplitude and vertical cutoff rigidity dependence can be described by a power law \citep{lockwood1991}.
Figure 8 (upper panel) shows a scatter plot of the cutoff rigidity \textit{versus} the FD amplitude from the real 
time Neutron Monitor Database (NMDB) data on 26 August 2015.
We fit the data using least square fit and estimate the power law index as  $\gamma = -0.43 \pm 0.09$, that is in 
the range reported previously for Forbush decreases associated with CMEs \citep{cane00}. 
Solid squares show the \textit{New-Tupi} observations. 
Thus, it is in agreement with the presently accepted physical scenario that indicates the self-similar 
nature of Forbush decreases in different rigidity regimes \citep{raghav2014}. 

Using the same approach and the same ground based detectors, we also investigated the rigidity dependence of the particle enhancement on 28 August 2015. 
The result is shown in Figure 8 (bottom panel), where the confidence level of the particle enhancement is plotted \textit{versus} the geomagnetic rigidity cutoff. 
The estimated  power law index appeared to be less steep, it is $\beta = -0.18 \pm 0.07$.

We have to mention here that while the neutron monitors  detect secondary baryons produced by particles (mostly galactic cosmic rays) colliding with nuclei in the atmosphere, the \textit{New-Tupi} telescopes register muons in the atmospheric cascade generated by the protons in the terrestrial atmosphere. Nevertheless, there are some similarities.
These detectors have similar energy threshold, close to $\sim 100$ MeV.
Also, it was shown that the correlation coefficient between the two detection systems is close to $\sim 0.8046$ \citep{kim12}.

\begin{figure}
\vspace{-0.0cm}
\centerline{\includegraphics[width=1\textwidth,clip=]{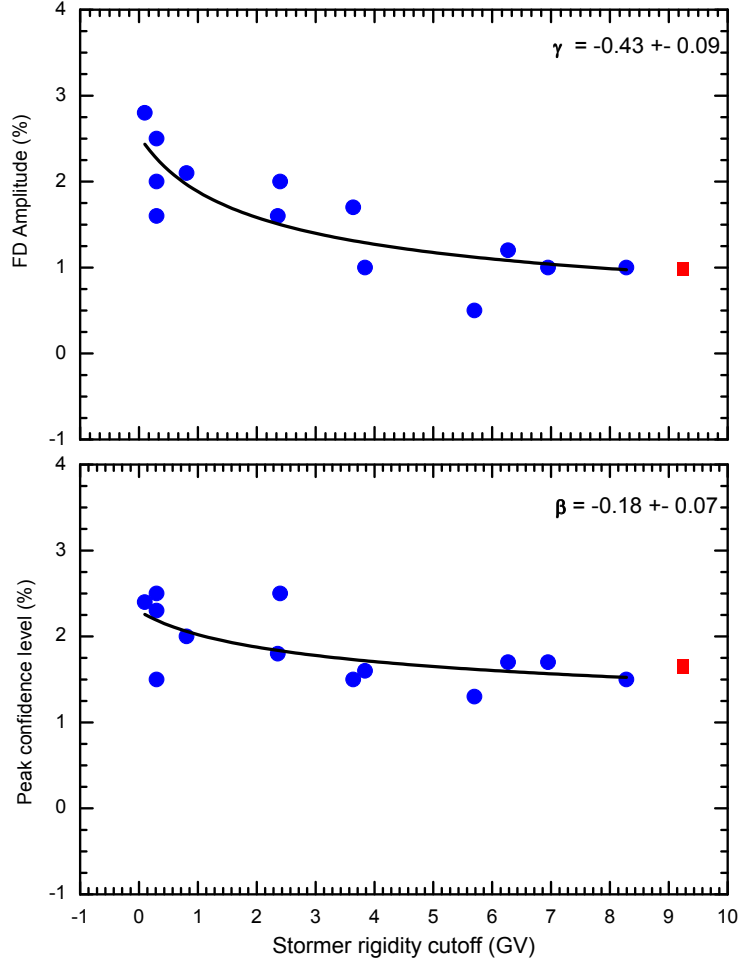}}
\vspace{-2.0cm}
\caption{Rigidity dependence of the FD amplitude on 26 August 2015  (top panel) and the peak confidence level (in percents) of 
the particle enhancement on 28 August 2015 (lower panel). In both cases the Real Time Neutron Monitor Database (NMDB) data (solid circles) have been used. Solid square shows the \textit{New-Tupi} telescopes observation.}
\label{fig8}
\end{figure}

\section{Conclusions}

We have analyzed the association between the muon flux variation at ground level, registered by the \textit{New-Tupi} muon 
telescopes and the geomagnetic storm of 25\,--\,29 August 2015 due to transient effects from the 22 August 2015 CME.
The \textit{New-Tupi} muon telescopes observed a Forbush decrease (FD) triggered by this geomagnetic storm, with onset on 26 August 2015.
These observations are studied in correlation with data obtained by space-borne detectors (ACE, GOES) and other ground-based experiments.
After the Earth crossed a heliospheric current sheet (HCS), a sudden increase in the particle flux was observed on 28 August 2015 by spacecrafts 
and ground level detectors in temporal coincidence with the impact of the high speed stream.
We have examined possible sources of the 28 August 2015 peak that was observed in a complex situation of the lingering effects of a CME, the HCS 
crossing, and the high speed stream impact.
Our result shows possible evidence of a prolonged energetic (up to GeV energies) particle injection within the Earth atmosphere system, driven by the HSS. 
However, a residual focusing effect of the HCS crossing and modulation of cosmic rays can not be excluded.
Further experimental observations are necessary to get better statistical constraints.

Despite the fact that both the FD and the particle enhancement show a small amplitude ($\sim 1\%$), we found that there is a high power 
variance obtained through the CWT spectra analysis.
Previous studies  of the high amplitude  FD (above 10\%)  showed  that common physical mechanisms, such as turbulence in 
the magnetic field associated with shock and ejecta, are characterized by the power law index. 
Here we found that the  low intensity FD amplitude and the confidence level of the particle enhancement in the different rigidity 
regimes also follow a power law behavior.
This can indicate the self-similar nature of  underlying formation mechanisms.
Better understanding can lead to the devising of new methods to detect their precursors. 

In addition, the lower limit of the integrated time fluence was obtained, as $1.4 \times 10^{-4} \rm erg\;cm^{-2}$. 
The injected energy in the upper atmosphere during $\sim 350$ min must be higher than this value. That could  have consequences for the atmospheric chemistry, for instance, the creation of NOx species may be enhanced and can lead to increased ozone depletion \citep{seppala07}. 
This topic also requires further study.
%CHANGE
%%%\cite{longden08}
%The energy of the inject particles is above  pion production ($E>$GeV), because muons are produced in the atmosphere. It is also expected a different chemical compositionthan the solar wind particles, because they are particles (ions) of the interplanetary space, accelerated by shock waves at the front edges of the transition zones between fast and slow moving solar wind.
 
\begin{acks}
This work is supported by the National Council for Research (CNPq) of Brazil, under Grant 306605/2009-0 and Fundacao de Amparo 
a Pesquisa do Estado do Rio de Janeiro (FAPERJ), under Grant 08458.009577/2011-81 and E-26/101.649/2011.
We express our gratitude to the ACE/MAG instrument team, the ACE Science Center, the NASA GOES team and the NOAA Space Weather 
Prediction Center (\url{www.swpc.noaa.gov}) for valuable information and for real time data. 
The Dst indices were provided by the World Data Center for Geomagnetism at Kyoto University, Japan (\url{wdc.kugi.kyoto-u.ac.jp}).
We acknowledge the NMDB database (\url{www.nmdb.eu}), founded under the European Union's FP7 programme (contract no. 213007) for providing data.
%CHANGE
The authors thank the anonymous referee for assistance, advice, very valuable and helpful suggestions.
The authors declare that they have no conflicts of interest.
\end{acks}

%%%%%%%%%%%%%%%%%%%%%%%%%%%%%%%%%%%%%%%%%%%%%%%%%%%
%% Sections
%
% \section{}%\label{s:?} 

%% Figure 
%
% \begin{figure} 
% \centerline{\includegraphics[width=0.5\textwidth,clip=]{<fig.eps>}}
% \caption{}%\label{fig:?}
% \end{figure}

%% Table
%
% \begin{table}
% \caption{}%\label{tbl:?}
% \begin{tabular}{}     
% \hline
% \multicolumn{2}{c}{<>}
% <data>
% \hline
% \end{tabular}
% \end{table}

%%%%%%%%%%%%%%%%%%%%%%%%%%%%%%%%%%%%%%%%%%%%%%%%%%%%%%%%%%%%%%%%%%%%%%%%%%%
%% Appendix
%
% \appendix   

%%%%%%%%%%%%%%%%%%%%%%%%%%%%%%%%%%%%%%%%%%%%%%%%%%%%%%%%%%%%%%%%%%%%%%%%%%%
%% Acknowledgements
%
% \begin{acks}
%
% \end{acks}

%%% %%%%%%%%%%%%%%%%%%%%%%%%%%%%%%%%%%%%%%%%%%%%%%%%%%%%%%%%%%%
%% Bibliography
%
% Using BibTeX
%
% \bibliographystyle{spr-mp-sola}
% \bibliography{<bib file>}  
%
% Without BibTeX 
% \begin{thebibliography}{}
% \bibitem[\protect\citeauthoryear{Author}{Year}]{key}
%   <bibliographical entry>
%
% \bibitem[\protect\citeauthoryear{}{}]{}
%   
%  
% \end{thebibliography}

\end{article} 
\end{document}